\documentclass{emulateapj}
\usepackage{graphicx}

\newcommand{\nh}{\overline{n}_{\rm H}}
\newcommand{\Nh}{N_{\rm H}}
\newcommand{\Nhtwo}{N_{\rm H_2}}
\newcommand{\calC}{\mathcal{C}}
\newcommand{\calR}{\mathcal{R}}
\newcommand{\dc}{\delta_{\rm C}}
\newcommand{\tcii}{T_{\rm CII}}
\newcommand{\msun}{M_{\odot}}
\newcommand{\ltsim}{\protect\raisebox{-0.5ex}{$\:\stackrel{\textstyle <}
	{\sim}\:$}}
\newcommand{\gtsim}{\protect\raisebox{-0.5ex}{$\:\stackrel{\textstyle >}
	{\sim}\:$}}

\begin{document}

\slugcomment{ApJ in press}

\title{Star Formation in Atomic Gas}

\author{Mark R.~Krumholz}
\affil{Department of Astronomy \& Astrophysics, University of California, Santa 
Cruz, CA 95064 USA}
\email{krumholz@ucolick.org}

\begin{abstract}
Observations of nearby galaxies have firmly established, over a broad range of galactic environments and metallicities, that star formation occurs exclusively in the molecular phase of the interstellar medium (ISM). Theoretical models show that this association results from the correlation between chemical phase, shielding, and temperature. Interstellar gas converts from atomic to molecular only in regions that are well shielded from interstellar ultraviolet (UV) photons, and since UV photons are also the dominant source of interstellar heating, only in these shielded regions does the gas become cold enough to be subject to Jeans instability. However, while the equilibrium temperature and chemical state of interstellar gas are well-correlated, the time scale required to reach chemical equilibrium is much longer than that required to reach thermal equilibrium, and both timescales are metallicity-dependent. Here I show that the difference in time scales implies that, at metallicities below a few percent of the Solar value, well-shielded gas will reach low temperatures and proceed to star formation before the bulk of it is able to convert from atomic to molecular. As a result, at extremely low metallicities, star formation will occur in a cold atomic phase of the ISM rather than a molecular phase. I calculate the observable consequences of this result for star formation in low metallicity galaxies, and I discuss how some current numerical models for H$_2$-regulated star-formation may need to be modified.
\end{abstract}


\section{Introduction}

In present day galaxies, star formation is very well-correlated with the molecular phase of the interstellar medium (ISM) \citep{wong02a, kennicutt07a, leroy08a, bigiel08a}. In contrast, in the inner parts of disks where there are significant molecular fractions, star formation correlates very poorly or not at all with the atomic ISM. At large galctocentric radii where the ISM becomes atomic-dominated star formation does begin to correlate with H~\textsc{i}, but this appears to be only because H$_2$ itself becomes correlated with H~\textsc{i}, and the H$_2$ forms stars in the same way regardless of where it is found within a galaxy \citep{bigiel10a, schruba11a}. Strong association between star formation and H$_2$ and a lack of association with H~\textsc{i} is also found down the lowest metallicity systems that have been measured, at roughly 20\% of Solar \citep{bolatto11a}. In summary, all available observational data indicates that star formation occurs only where the hydrogen in the ISM has converted to H$_2$.

Theoretical models have explained these observations as resulting from a correlation between chemistry and temperature \citep{schaye04a, krumholz11b, glover12a}. Molecular hydrogen is not an important coolant in modern-day galaxies, and while carbon monoxide (which forms only when it is catalyzed by H$_2$ -- \citealt{van-dishoeck86a}) is, the C~\textsc{ii} found in H~\textsc{i} regions is almost as effective. However, H$_2$ is an excellent proxy for the presence of cold gas because both are sensitive to destruction by UV photons, which photodissociate H$_2$ and increase the temperature through the grain photoelectric effect. As a result, both H$_2$ and low temperature gas are found only in regions of high extinction where the UV photon density is far below its mean value in the ISM, and, conversely, any region that where the photodissociation rate is high enough to convert the bulk of the ISM to H~\textsc{i} is also likely to be warm. Since low temperatures that remove thermal pressure support are a prerequisite for collapse into stars, this correlation between temperature and chemical state in turn induces a correlation between star formation and chemical state.

However, the correlation between H$_2$ and star formation must break down at sufficiently low metallicities. Before the first stars formed in the universe, and for a short time thereafter, there were no or very few heavy elements. As a result, forming H$_2$ was extremely difficult due to a lack of dust grain surfaces to catalyze the ${\rm H~\textsc{i}} \rightarrow {\rm H}_2$ reaction. Theoretical models of star formation in such environments indicate that H$_2$ fractions remain extremely small until the density rises so high ($\gtsim 10^9$ cm$^{-3}$) that H$_2$ can form via three-body reactions \citep{palla83a, lepp84a, ahn07a, omukai10a}. The underlying physical basis for this result is a disconnect of timescales: the equilibrium chemical state the gas would reach after a very long time would be H$_2$-dominated, but the cooling and star formation times are short enough that the gas does not reach equilibrium before collapsing into a star.

While this result has been known for zero and extremely low metallicity systems for some time, the relationship between chemical state and star formation in intermediate metallicity regime, for which observations are at least in principle possible in the local universe, has received fairly little attention. \citet{omukai10a} consider the chemical evolution of collapsing gas cores with metallicities from 0 to Solar, and investigate under what circumstances they can form H$_2$. However, because their calculation starts with unstable, collapsing cores, it does not address the question of in what phase of the ISM one expects to find such collapsing regions in the first place, which is the central problem for understanding the observed galactic-scale correlation between ISM chemical state and star formation. \citet{glover12b} simulate the non-equilibrium chemical and thermal behavior of clouds with metallicities from 1\% of Solar to Solar. They find that the bulk of the cloud material converts to H$_2$ before star formation in the high metallicity clouds but not in the lowest metallicity ones, indicating that the star formation - H$_2$ correlation should begin to break down at metallicities observable in nearby galaxies. However, given the computational cost of their simulations, they are able to explore a very limited number of cases, and it is unclear how general their results might be.

The goal of this paper is to go beyond the studies of \citet{omukai10a} and \citet{glover12b} by deriving general results about the correlation between chemical state and star formation over a wide range of environments and metallicities. I do not perform detailed simulations, such as those of \citeauthor{omukai10a} and \citeauthor{glover12b}, for every case. Instead, I rely on fairly simple models that can be integrated semi-analytically. The benefit of this approach is that it is the only way to survey a large parameter space, and thereby to answer the central questions with which I am concerned: under what conditions do we expect the correlation between star formation and H$_2$ to break down? When such a breakdown occurs, what is the governing physical mechanism that causes it? What are the resulting observational signatures? What are the implications of this breakdown for the models of star formation commonly adopted in studies of galaxy formation? In the remainder of this paper, I seek to answer these questions.

\section{Model}

Consider spatially uniform gas characterized by a mean number density of H nuclei $\nh$, column density of hydrogen nuclei $\Nh$, metallicity $Z'$ relative to Solar, and temperature $T$. A fraction $f_{\rm H_2}$ of the H nuclei are locked in H$_2$ molecules. It is generally more convenient to characterize models by values of the visual extinction $A_V$ instead of $\Nh$. These two are related by $A_V/\Nh \approx 4.0 \times 10^{-22} Z'$ mag cm$^2$, with the normalization chosen as a compromise between the values for Milky Way extinction and the extinction curves of the Large and Small Magellanic Clouds adjusted to Milky Way metallicity.\footnote{The value of $A_V/\Nh$, and all other parameters in the following discussion, are taken from \citet{krumholz11b} unless stated otherwise; arguments for these choices are given in that paper. Atomic data (Einstein coefficients, collision rates) are all taken from \citet{schoier05a}, and abundances from \citet{draine11a}.}

\subsection{Timescale Estimates}

We are interested in following the behavior of initially warm, atomic gas, and considering whether it will be able to cool to temperatures low enough to allow star formation, and how its chemical state will evolve as it does so. Before computing detailed evolutionary histories, it is helpful first to make a rough estimate of the timescales involved. In interstellar gas that is dense enough to be a candidate for star formation, but that is not yet molecular or forming stars, the dominant cooling process is emission in the [C~\textsc{ii}] 158 $\mu$m line, which removes energy at a rate
\begin{equation}
\Lambda_{\rm CII} \approx k_{\rm CII-H} \dc k_B \tcii \calC \nh
\end{equation}
per H atom, where $k_{\rm CII-H}\approx 8\times 10^{-10} e^{-\tcii/T}$ cm$^3$ s$^{-1}$ is the rate coefficient for collisional excitation of C~\textsc{ii} by H atoms, $\dc\approx 1.1\times 10^{-4} Z'$ is the gas phase carbon abundance, $\tcii=91$ K is the energy of the excited C~\textsc{ii} level over $k_B$, and $\calC = \langle n_{\rm H}^2 \rangle / \nh^2$ is a clumping factor that accounts for clumping of the medium on size scales below that on which we are computing the average. This expression assumes that C~\textsc{ii} collisional excitation is dominated by H rather than by free electrons, that the gas is optically thin, and that the density is far below the critical density for the line; I show below that all these assumptions are valid. The time required for the gas to reach thermal equilibrium is of order
\begin{eqnarray}
t_{\rm therm} & \equiv & \frac{k_B T}{\Lambda_{\rm CII}} = 
\frac{T}{k_{\rm CII-H} \delta_{\rm C} \tcii \calC \nh} 
\nonumber \\
& = & 0.036 \left(\frac{T}{\tcii}\right)e^{\tcii/T} Z'^{-1} \calC_{1}^{-1} n_0^{-1}\mbox{ Myr},
\end{eqnarray}
where $\calC_{1} = \calC/10$ and $n_0 = \nh/1$ cm$^{-3}$. Note that the value of $\calC$ will depend on the size scale over which the average density is defined; the fiducial value $\calC = 10$ is intermediate between the values $\calC \approx 2$ and $\calC \approx 30$ that numerical experiments indicate are best for $\sim 10$ pc and $\sim 100$ pc scale, respectively \citep{gnedin09a, mac-low12a}.

Conversion of the gas from atomic to molecular form occurs primarily on the surface of dust grains down to metallicities as low as $\sim 10^{-5}$ of Solar \citep{omukai10a}. This process occurs at a rate per H atom $\nh \calR \calC$, where $\mathcal{R}\approx 3\times 10^{-17} Z'$ cm$^3$ s$^{-1}$ is the rate coefficient for H$_2$ formation on grain surfaces \citep{wolfire08a}. The associated timescale for conversion of the gas to molecular form is
\begin{equation}
t_{\rm chem} \equiv \frac{1}{\nh \calR \calC} = 105 Z'^{-1} n_0^{-1} \calC_{1}^{-1} \mbox{ Myr}.
\end{equation}
The ratio of the two timescales is
\begin{equation}
\frac{t_{\rm chem}}{t_{\rm therm}} = \frac{k_{\rm CII-H} \delta_{\rm C} \tcii}{\calR T} = 2900 \left(\frac{\tcii}{T}\right) e^{-\tcii/T},
\end{equation}
indicating that the gas will reach thermal equilibrium vastly before it reaches chemical equilibrium.

This difference in timescale is only important if the cooling of gas toward thermal equilibrium is followed by star formation on a timescale that is too short for the conversion of atomic to molecular gas to keep up. It is therefore helpful to consider a third timescale: the free-fall time
\begin{equation}
t_{\rm ff} = \sqrt{\frac{3\pi}{32 G \nh\mu_{\rm H} m_{\rm H}}} = 43 n_0^{-1/2}\mbox{ Myr},
\end{equation}
where $\mu_{\rm H} \approx 1.4$ is the mean mass per H nucleus in units of the hydrogen mass $m_{\rm H}$. This is the timescale over which star formation should begin once gas is gravitationally unstable. The time for which a cloud survives after the onset of star formation is significantly uncertain, but even the longest modern estimates are $\sim 10 t_{\rm ff}$, while some are as short as $\sim 1 t_{\rm ff}$ \citep{elmegreen00a, tan06a, kawamura09a, goldbaum11a}. Comparing this timescale to the two previously computed gives
\begin{eqnarray}
\frac{t_{\rm therm}}{t_{\rm ff}} & = & 8.3\times 10^{-4} \left(\frac{T}{\tcii}\right)e^{\tcii/T} Z'^{-1} \calC_{1}^{-1} n_0^{-1/2}
\\
\frac{t_{\rm chem}}{t_{\rm ff}} & = & 2.4 Z'^{-1} \calC_{1}^{-1} n_0^{-1/2}.
\end{eqnarray}
Thus we see that the thermal timescale will be smaller than the free-fall timescale down to extremely low metallicities for reasonable ISM densities and temperatures, but the same cannot be said of the chemical timescale. For example, at $n_0 = 100$, $T = 1000$ K, $Z' = 10^{-3}$, and $\calC_1=1$, the above equation gives $t_{\rm therm}/t_{\rm ff} = 1.0$, while $t_{\rm chem}/t_{\rm ff} = 240$. A cloud with these properties could cool and proceed to star formation on a free-fall timescale without difficulty, but would not build up a substantial amount of H$_2$ until more than 100 free-fall times. If such a cloud were anything like the observed star-forming clouds in the Milky Way, it would likely have been destroyed by stellar feedback well before this point. Note that individual overdense regions within the cloud in the process of collapsing to stars would still convert to H$_2$, since $t_{\rm chem}/t_{\rm ff}$ is a decreasing function of density, dramatically so once three-body reactions begin to occur; however, since the star formation efficiency is low, the non-star-forming bulk of the cloud material would not, and thus the amount of H$_2$ present per unit star formation would be greatly reduced.

\subsection{Chemical and Thermal Evolution Models}

The argument above is based on simple timescale estimates. To check whether the result is robust, it is necessary to construct more sophisticated cooling and chemistry models. Below I describe a more detailed model, and how it may be evaluated numerically to follow the thermal and chemical behavior of a cloud.

\subsubsection{Chemical Evolution}

The H~\textsc{i} to H$_2$ transition is governed by two main processes: formation of H$_2$ on the surfaces of dust grains and destruction of H$_2$ by photodissociation. The former occurs at a rate per H atom $n_{\rm H} \calR$, where $n_{\rm H}$ is the local (rather than average) number density and $\calR$ is the metallicity-dependent rate coefficient given in the main text. In a region of mean density $\nh$, the mean rate per H nucleus at which H~\textsc{i} converts to H$_2$ is simply the number density-weighted average of the rate given above, which is $\calC \nh \calR$. The photodissociation rate per H$_2$ molcule is
\begin{equation}
\zeta_{\rm diss} = \zeta_{\rm diss,0} e^{-\sigma_d \Nh} f_{\rm shield}(\Nhtwo),
\end{equation}
where $\zeta_{\rm diss,0}\approx 5\times 10^{-11}$ s$^{-1}$ is the dissociation rate for unshielded gas \citep{draine96a}, $\sigma_d\approx 10^{-21} Z'$ cm$^{-2}$ is the dust cross section per H nucleus for Lyman-Werner band photons, $\Nh$ is the column density of H nuclei, $\Nhtwo$ is the column density of H$_2$ molecules, and $f_{\rm shield}(\Nhtwo)$ is the shielding function that describes H$_2$ self-shielding. For the latter quantity, I use the approximate form of \citet{draine96a},
\begin{equation}
f_{\rm shield} \approx \frac{0.965}{(1+0.1x/b_6)^2} + \frac{0.035}{(1+x)^{0.5}} e^{-8.5\times 10^{-4}(1+x)^{0.5}},
\end{equation}
where $x=\Nhtwo/5\times 10^{14}$ cm$^{-2}$ and $b_6$ is the Doppler parameter for the gas in units of $10^6$ cm s$^{-1}$; I use $b_6 = 0.71$, corresponding to a velocity dispersion of a 5 km s$^{-1}$, roughly that observed in molecular clouds in nearby galaxies, but the results are quite insensitive to this choice. The value of $\zeta_{\rm diss,0}$ is that appropriate for the Milky Way's radiation field, and this value will vary from galaxy to galaxy and within galaxies. However, I show below that the dependence of the results on this choice is also quite weak, because of the exponential dependence of the dissociation rate on column density: a relatively large change in $\zeta_0$ can be compensated for by a far smaller change in $\Nh$ or $\Nhtwo$ \citep[also see][]{krumholz11b}.

Given these processes, the rate of change of the H$_2$ fraction is given by
\begin{equation}
\label{eq:dfh2dt}
\frac{d}{dt}f_{\rm H_2} = (1-f_{\rm H_2}) \nh \calC \calR - f_{\rm H_2} \zeta_{\rm diss}(\Nh,\Nhtwo).
\end{equation}
Consistent with the simple uniform cloud assumption, I adopt $\Nhtwo = f_{\rm H_2} \Nh/2$. Note that there is no factor of 2 or $1/2$ in the second term due to a cancellation: there is one H$_2$ molecule per two H nuclei bound as H$_2$ (multiplying by a factor of $1/2$), but each dissociation generates two free H nuclei (multiplying by a factor of two).

\subsubsection{Thermal Evolution}

The thermal evolution depends on heating and cooling processes. For heating, the dominant mechanisms are cosmic ray heating and the grain photoelectric effect. The photoelectric heating rate per H nucleus is
\begin{equation}
\Gamma_{\rm PE} = \Gamma_{\rm PE,0} Z' e^{-\sigma_d \Nh}
\end{equation}
where $\Gamma_{\rm PE,0} \approx  4\times 10^{-26}$ erg s$^{-1}$ is the grain photoelectric heating rate in free space, and the numerical value is for a Milky Way radiation field. Note that, since photoelectric heating is dominated by photons with energies similar to the Lyman-Werner bands, the dust cross section $\sigma_d$ here is the same as that used in the H$_2$ formation calculation. The heating rate per H nucleus from cosmic rays is
\begin{equation}
\Gamma_{\rm CR} = \zeta_{\rm CR} q_{\rm CR}
\end{equation}
where $\zeta_{\rm CR}$ is the primary cosmic ray ionization rate and $q_{\rm CR}$ is the energy added per primary cosmic ray ionization. For atomic gas, $q_{\rm CR} \approx 6.5$ eV \citep{dalgarno72a}. The observed primary cosmic ray ionization rate in the Milky Way varies sharply between diffuse sightlines and dark clouds; the former show $\zeta_{\rm CR} \approx 3\times 10^{-16}$ s$^{-1}$, with roughly a dex dispersion, while those in dark clouds are an order of magnitude lower, $\zeta_{\rm CR} \approx 2 \times 10^{-17}$ s$^{-1}$ \citep{wolfire10a, neufeld10a, indriolo12a}. Since cosmic ray heating is only significant compared to photoelectric heating in dark clouds, it seems more reasonable to adopt the latter as a fiducial value, although I verify below that this choice does not significantly affect the results. Observations also indicate that cosmic ray ionization rates vary roughly linearly with galactic star formation rates \citep{abdo10a}. For these reasons I adopt $\zeta_{\rm CR} = 2\times 10^{-17} Z'$ s$^{-1}$ as a fiducial value; the metallicity scaling is a very rough way of accounting for the lower cosmic ray flux in galaxies with lower metallicities, masses, and star formation rates \citep{krumholz11b}.

Cooling for interstellar gas with temperatures $\ll 10^4$ K is dominated by the 158 $\mu$m fine structure line of C~\textsc{ii} and the $63$ and $145$ $\mu$m fine structure lines of O~\textsc{i}. The critical densities for these transitions are approximately $4\times 10^3$, $6\times 10^4$, and $3\times 10^5$ cm$^{-3}$, respectively assuming the dominant collision partner is H (see below). These critical densities are significantly higher than the densest cases I consider, and so I neglect collisional de-excitation compared to radiative de-excitation. Assuming all cloud atoms are in the ground state, the line-center optical depth of a cloud to photons emitted in one of these lines is
\begin{equation}
\tau_0 = \frac{g_u}{g_\ell} \frac{A_{u\ell} \lambda_{u\ell}^3}{8 \pi^{3/2}b} \delta_X \Nh = [0.11, 0.38, 0.51] Z' A_{V,0} b_6^{-1}
\end{equation}
where $g_u$ and $g_{\ell}$ are the degeneracies of the upper and lower states, $A_{u\ell}$ and $\lambda_{u\ell}$ are the Einstein $A$ and wavelength for the transition, $b$ is the Doppler parameter, $\delta_X$ is the abundance relative to hydrogen for the element in question, and $A_{V,0} = A_V/1$ mag. The three numbers given in square brackets are the numerical results for the lines [C~\textsc{ii}] 158 $\mu$m, [O~\textsc{i}] 63 $\mu$m, and [O~\textsc{i}] 145 $\mu$m, respectively, using abundances $\delta_{\rm C} = 1.1\times 10^{-4} Z'$ and $\delta_{\rm O} = 5.0\times 10^{-4} Z'$. This implies that, for the great majority of the cases I consider, optical depths effects will have at most a marginal effect on the cooling rate and can thus be neglected. For optically thin cooling at densities well below the critical density, the radiative cooling rate per H nucleus is simply
\begin{equation}
\Lambda_{\rm line} = \sum_i \frac{g_{u,i}}{g_{\ell,i}} \delta_i E_{u,i} e^{-E_{u,i}/k_B T} \calC (k_{i-\rm H} \nh + k_{i-e} \overline{n}_e),
\end{equation}
where the sum runs over the three upper states for the cooling lines (the $^2P_{3/2}^o$ state of C~\textsc{ii} and the $^3P_1$ and $^3P_0$ states of O~\textsc{i}), $g_{u,i}$ and $g_{\ell,i}$ are the degeneracies of the upper states and the corresponding ground states, $E_{u,i}$ is the energy of the upper state relative to ground, $\delta_i$ is the abundance of the relevant species,  $\overline{n}_e$ is the free electron density, and $k_{i-\rm H}$ and $k_{i-e}$ are the rate coefficients for collisional de-excitation of the level by H and by free electrons, respectively. The clumping factor $\calC$ appears for the same reason as for H$_2$ formation on dust grains: these are collisional processes whose rates vary as the square of the local volume density. Obviously once H$_2$ becomes dominant over H~\textsc{i} one should consider collisions with H$_2$ as well, but since conversion to H$_2$ only happens long after the gas has reached thermal equilibrium, I ignore this complication.

I take the free electron density to be $\overline{n}_e = \delta_{\rm C} \nh$, which assumes that singly ionized carbon is the dominant source of free electrons; with this choice, excitation by H generally dominates. At low column densities, free electrons produced by ionization of hydrogen by soft x-rays outnumber those coming from carbon, but in regions of column density $\Nh \gtsim 10^{20}$ cm$^{-2}$ this source of electrons becomes subdominant \citep{wolfire03a}. Since most of the cases with which I am concerned are in this regime, I adopt the carbon-dominated limit.

Combining all heating and cooling processes, the temperature evolution obeys
\begin{eqnarray}
\lefteqn{\frac{3}{2}k_B \frac{d}{dt} T = \Gamma_{\rm PE,0} Z' e^{-\sigma_d \Nh} + \zeta_{\rm CR} q_{\rm CR}}
\nonumber\\
& & {} - \sum_i \frac{g_{u,i}}{g_{\ell,i}} (k_{i-\rm H} + \delta_{\rm C} k_{i-e}) \delta_i E_{u,i} e^{-E_{u,i}/k_B T} \calC \nh.\quad
\label{eq:dTdt}
\end{eqnarray}

\section{Model Results}

\subsection{Fiducial Parameters}

In order to survey parameter space, I consider a grid of model clouds of density $\nh = 10^0 - 10^3$ cm$^{-3}$ in steps of 0.1 dex, extinction $A_V = 10^{-2} - 10^1$ mag in steps of 0.1 dex, and metallicity $Z' = 10^{-4} - 10^0$ in steps of 0.05 dex, using the fiducial values for radiation and cosmic rays specified above. All model clouds start with $f_{\rm H_2} = 0$ and $T = 1000$ K.\footnote{Obviously some of these initial conditions (e.g.~clouds with $\nh=10^3$ cm$^{-3}$ and $T=1000$ K) are unlikely to be found in real galaxies, but such models constitute a small fraction of the parameter space, and the goal of this study is to perform a broad survey rather than trying to focus on particular assumed sets of ``reasonable" initial conditions in galaxies where the actual conditions are poorly determined.}  I integrate each model for 15 Gyr and record the properties at $t=t_{\rm ff}$, $t=10t_{\rm ff}$, and at $t=15$ Gyr, with the latter representing the equilibrium state attained after long times; obviously this is longer than the age of the Universe, but these models are useful to help build intuition for the importance of non-equilibirum effects. From the temperatures produced in the models, I compute the Bonnor-Ebert mass $M_{\rm BE} = 1.18 c_s^3/\sqrt{G^3 \mu_{\rm H} m_{\rm H} n}$, where $c_s =\sqrt{k_B T/\mu m_{\rm H}}$ is the sound speed and $\mu \approx 1.3$ is the mean mass per particle (as opposed to the mean mass per H nucleus $\mu_{\rm H}$). Values of $M_{\rm BE}$ should serve a rough proxies for where star formation can occur, with values much larger than the mass of any star indicating little star formation, and small values indicating star formation.

\begin{figure}
\plotone{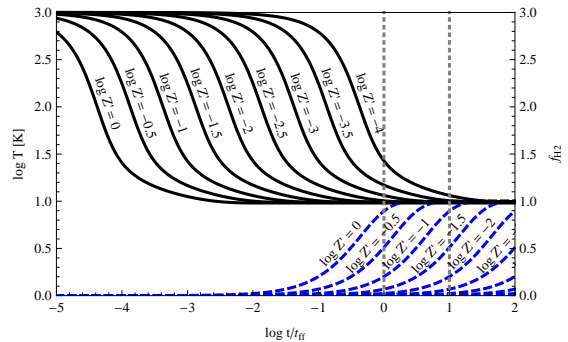}
\caption{
\label{fig:tchemevol}
Time evolution of the temperature (black solid lines) and H$_2$ fraction (blue dashed lines) for a cloud of initial temperature $T=1000$ K, density $n_0 = 30$ (clumping factor $\calC_1=1$), and extinction $A_V = 2$ mag, initially composed of pure H~\textsc{i}. Gray dotted vertical lines indicate times of $t=t_{\rm ff}$ and $t=10t_{\rm ff}$. Different lines are for metallicities of $\log Z' = -4 - 0$ in steps of 0.5, as indicated.
}
\end{figure}

Figure \ref{fig:tchemevol} shows the thermal and chemical evolution of some example clouds drawn from the model grid. The figure is consistent with the qualitative timescale estimates above: gas at an initially high, non-equilibrium temperature will cool to a thermal equilibrium temperature of order 10 K and thus proceed to star formation in less than a free-fall time, even for metallicities as low as $\log Z' \approx -4$. On the other hand, at metallicities below $\log Z' = -1$ the gas will be less than half converted to molecules at one free-fall time, and at metallicities of $\log Z' = -2$ or less the gas will not reach 50\% molecular until more than $10t_{\rm ff}$. This result is consistent with numerical experiments in full cosmological simulations which show that equilibrium models of the H$_2$ fraction begin to fail due to non-equilibirum effects at metallicites below $\log Z'\approx -2$ \citep{krumholz11a}.

\begin{figure}
\plotone{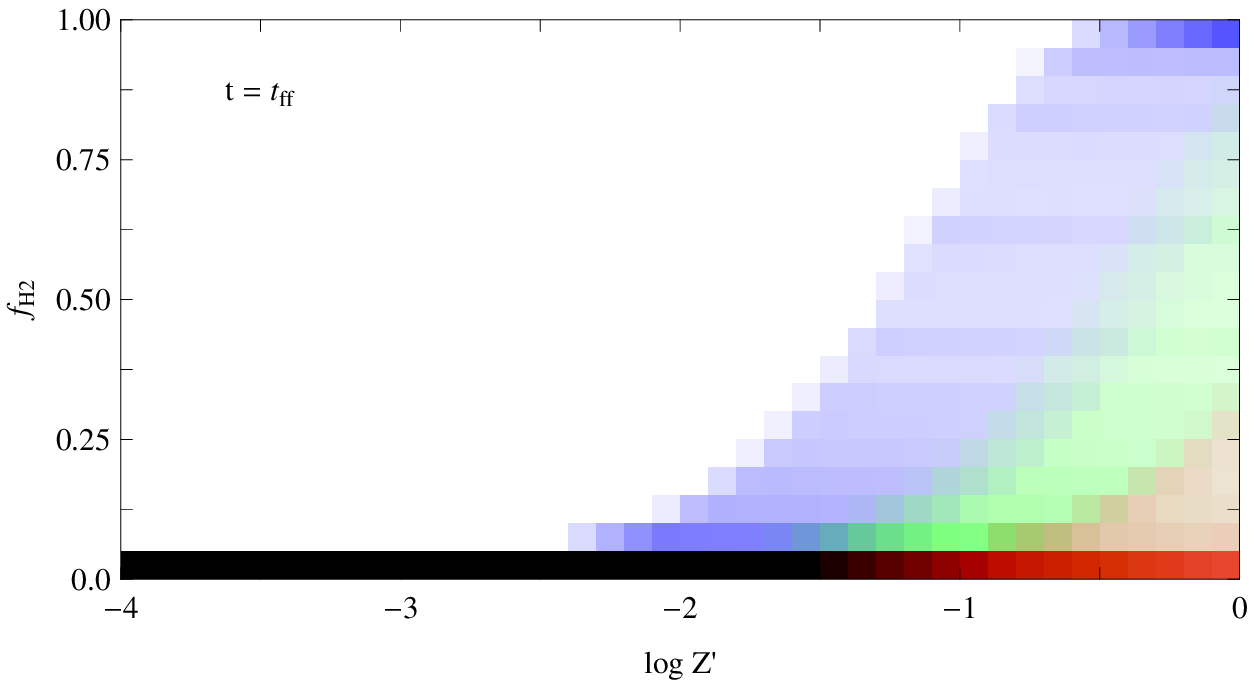}

\plotone{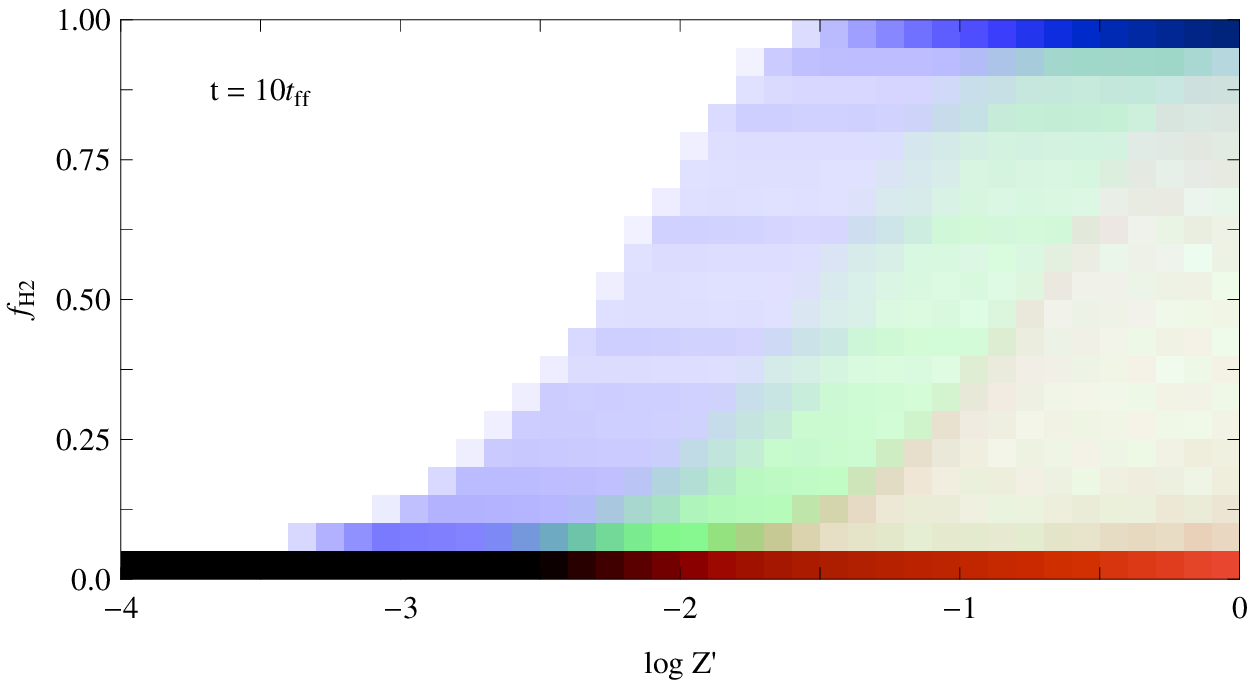}

\plotone{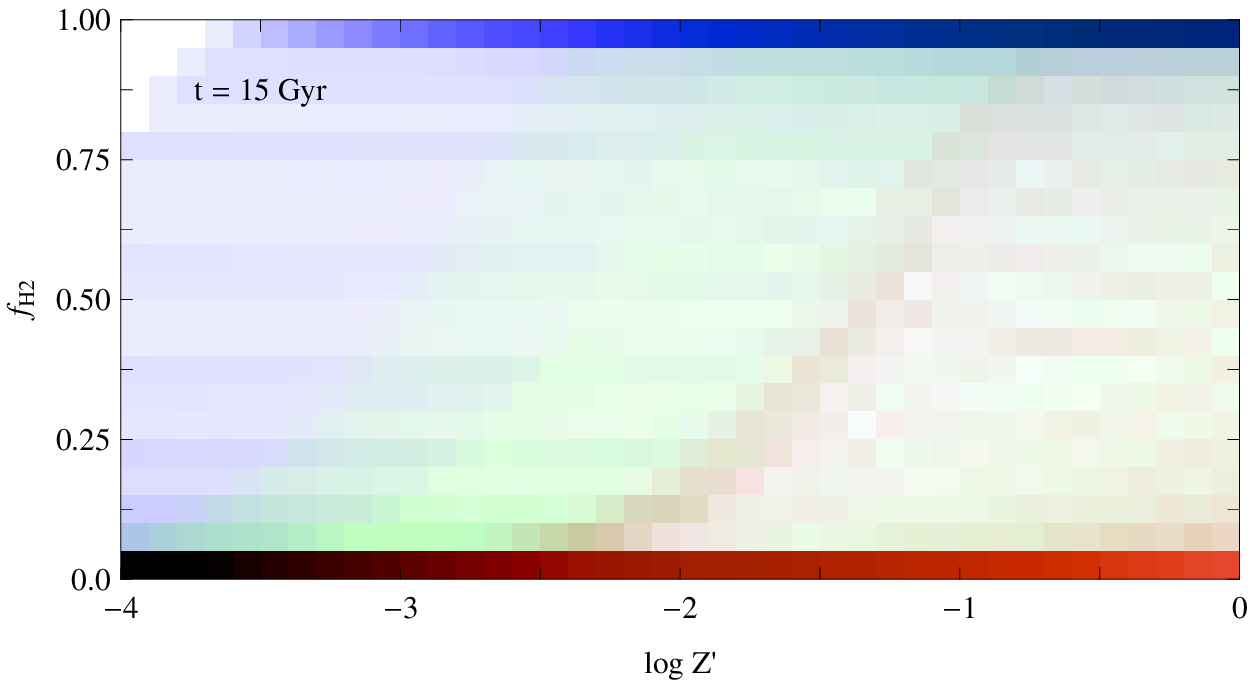}
\caption{
\label{fig:fH2Zgrid}
H$_2$ fraction versus metallicity for model clouds, with blue indicating star-forming clouds, red indicating non-star-forming clouds, and green indicating intermediate values. For each model in a grid covering the parameter range $\log Z' = -4 - 0$, $\log n_0 = 0 - 3$, $\log [A_V/{\rm mag}] = -2 - 1$, I compute the Bonnor-Ebert mass $M_{\rm BE}$ and the H$_2$ fraction at times $t = t_{\rm ff}$ (first panel), $t=10 t_{\rm ff}$ (second panel), and $t=15$ Gyr (third panel). Blue pixels indicate the locus in $\log Z'$, $f_{\rm H_2}$ corresponding to models for which $M_{\rm BE} < 100$ $\msun$, indicating the star formation in the cloud is likely. Red pixels indicate the locus of models with $M_{\rm BE} > 1000$ $\msun$ indicating star formation is unlikely; green pixels show intermediate values of $M_{\rm BE}$. For each color, brightness indicates what fraction of the models at that value of $Z'$ have a given H$_2$ fraction, with white representing none of the models, and solid blue, red, or green representing all of them. 
}
\end{figure}

Figure \ref{fig:fH2Zgrid} shows the H$_2$ fraction as a function of metallicity for star-forming, non-star-forming, and intermediate clouds at $t=t_{\rm ff}$, $t=10 t_{\rm ff}$, and $t=15$ Gyr (long enough that nearly all models have reached chemical equilibrium). In clouds that are very old and thus have reached equilibrium, the figure shows that star-forming clouds (those with low $M_{\rm BE}$, indicated in blue) lie almost exclusively at high H$_2$ fractions, and non-star-forming ones (those with high $M_{\rm BE}$, indicated in red) almost exclusively at low H$_2$ fractions, consistent with earlier work indicating a strong correlation between equilibrium gas temperature and chemical state \citep{krumholz11b, glover12a}. Out of equilibrium, the figure indicates that the correlation between low M$_{\rm BE}$ and high molecular fraction continues to hold for high metallicities. At lower metallicities, however, all models are displaced to smaller H$_2$ fractions, and at metallicities $\log Z' \ltsim -2$ star-forming clouds are likely to have H$_2$ fractions well below unity even at $t=10t_{\rm ff}$. The implication is that star-formation will be complete before the gas is significantly converted into H$_2$. The precise transition metallicity below which equilibrium is not achieved will depend on the value of the clumping factor and the timescale for which star-forming clouds typically survive.

\subsection{Sensitivity to Parameter Choices}

\begin{figure}
\plotone{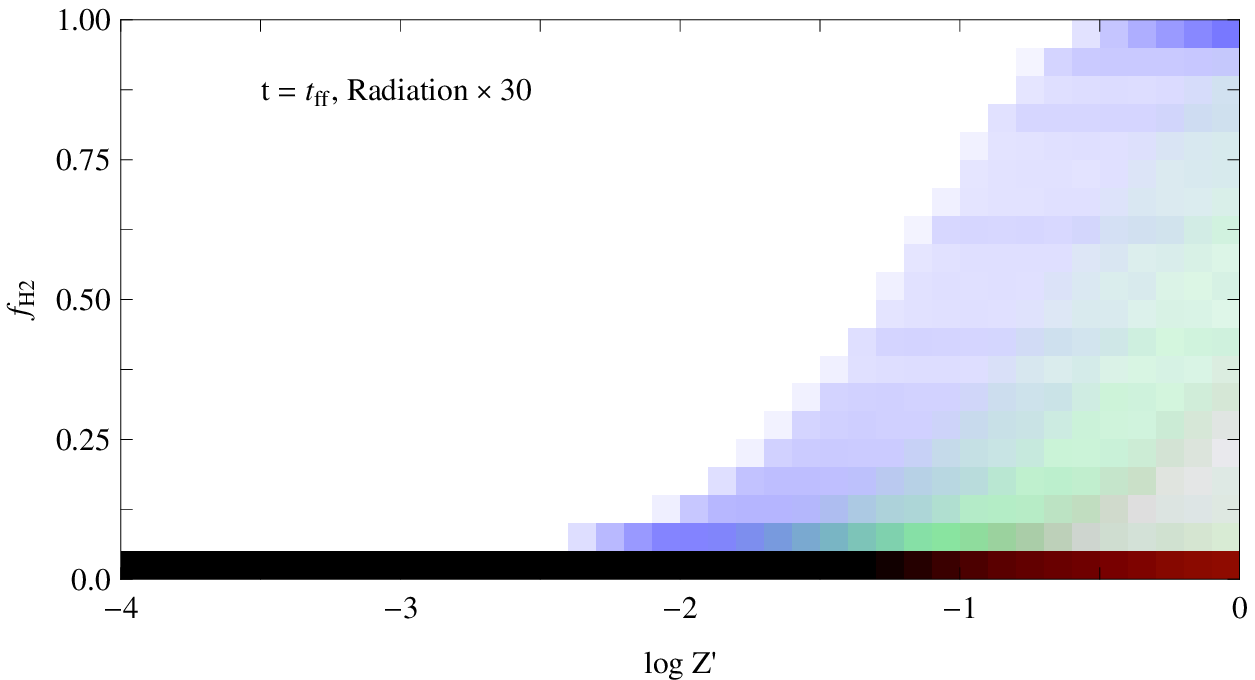}

\plotone{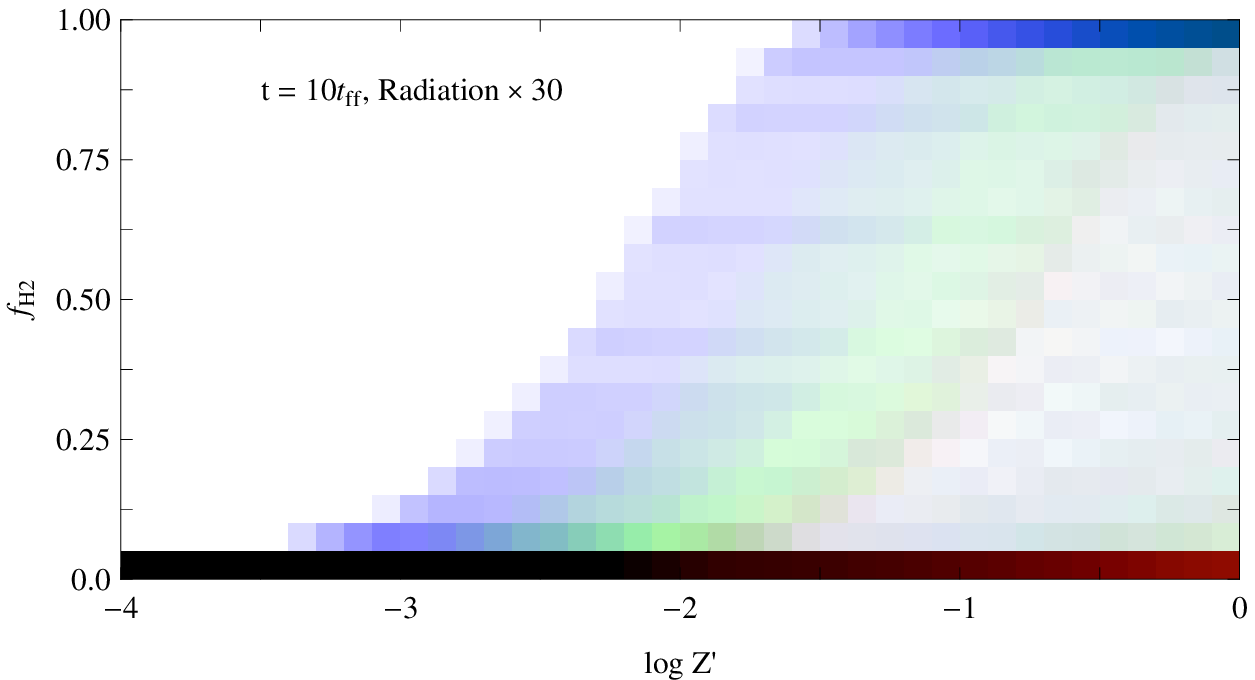}

\plotone{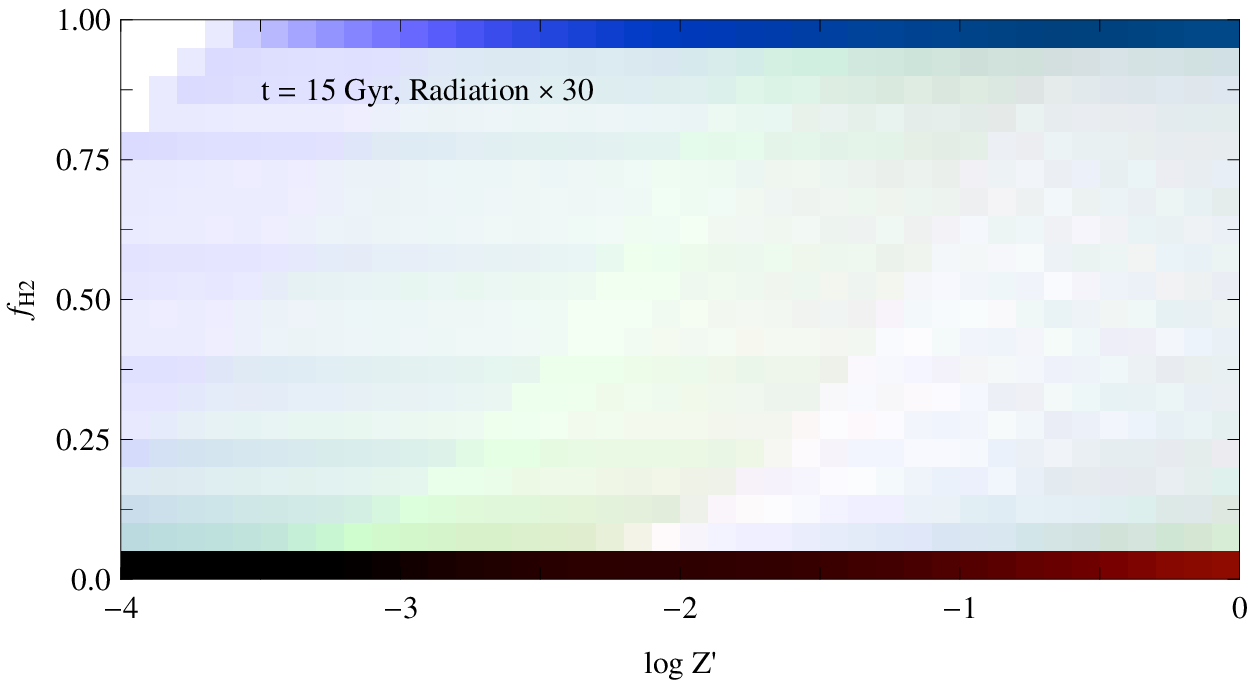}
\caption{
\label{fig:fH2Zgridhi}
Same as Figure \ref{fig:fH2Zgrid}, but computed for a UV radiation field 30 times stronger than the fiducial value.
}
\end{figure}

\begin{figure}
\plotone{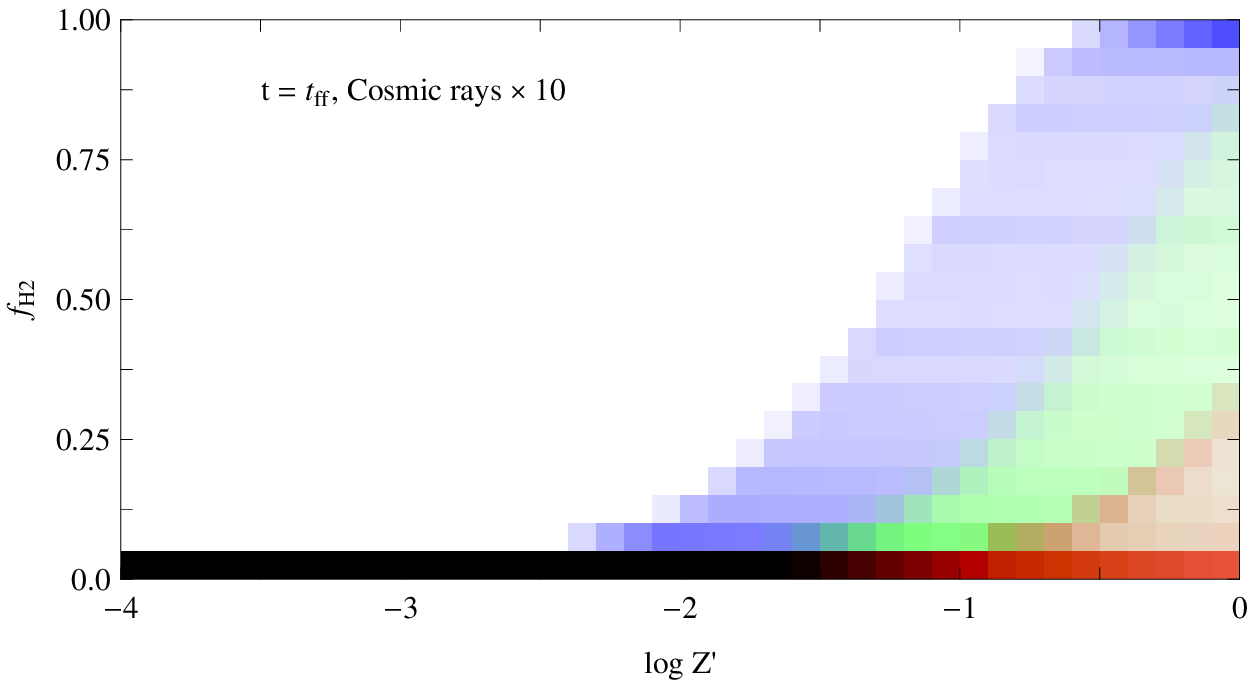}

\plotone{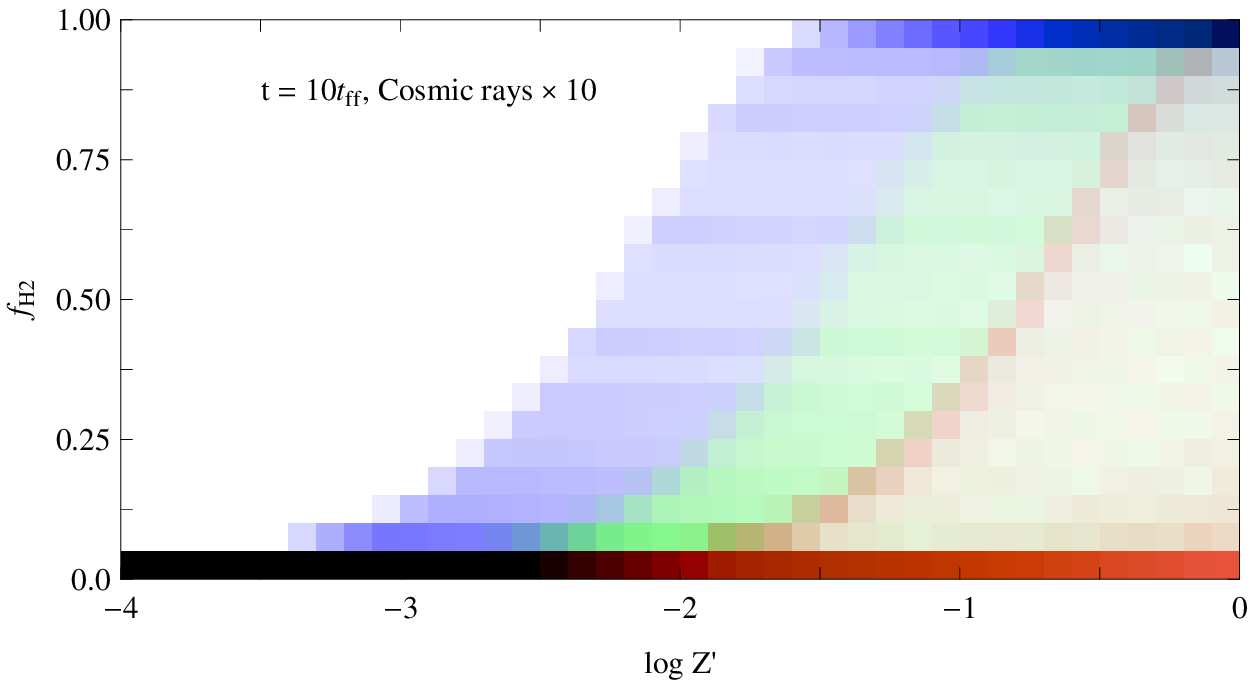}

\plotone{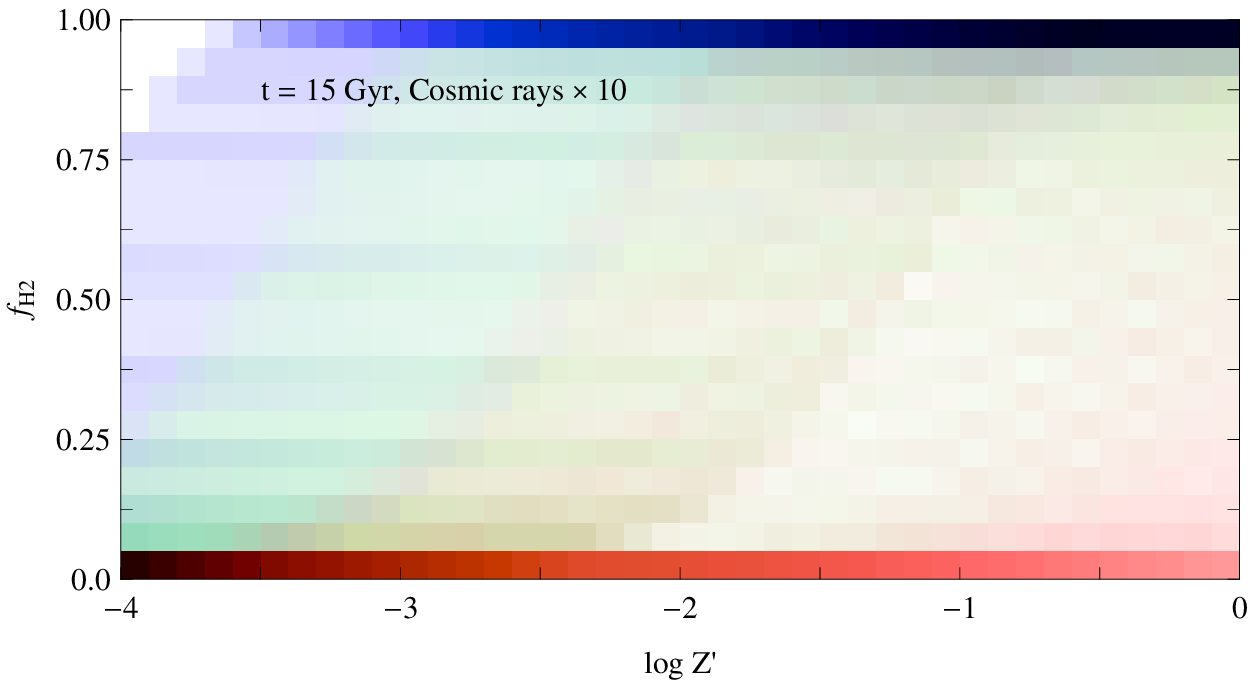}
\caption{
\label{fig:fH2Zgridhicr}
Same as Figure \ref{fig:fH2Zgrid}, but computed for cosmic ray ionization rate 10 times larger than the fiducial value.
}
\end{figure}

\begin{figure}
\plotone{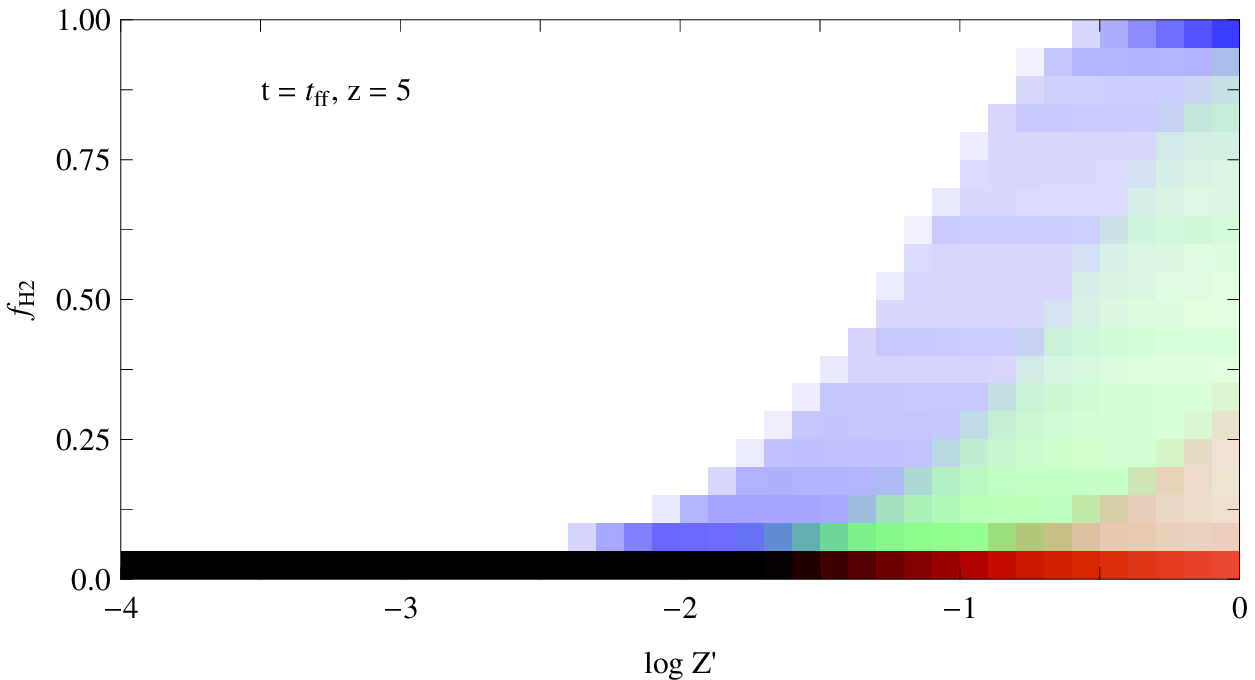}

\plotone{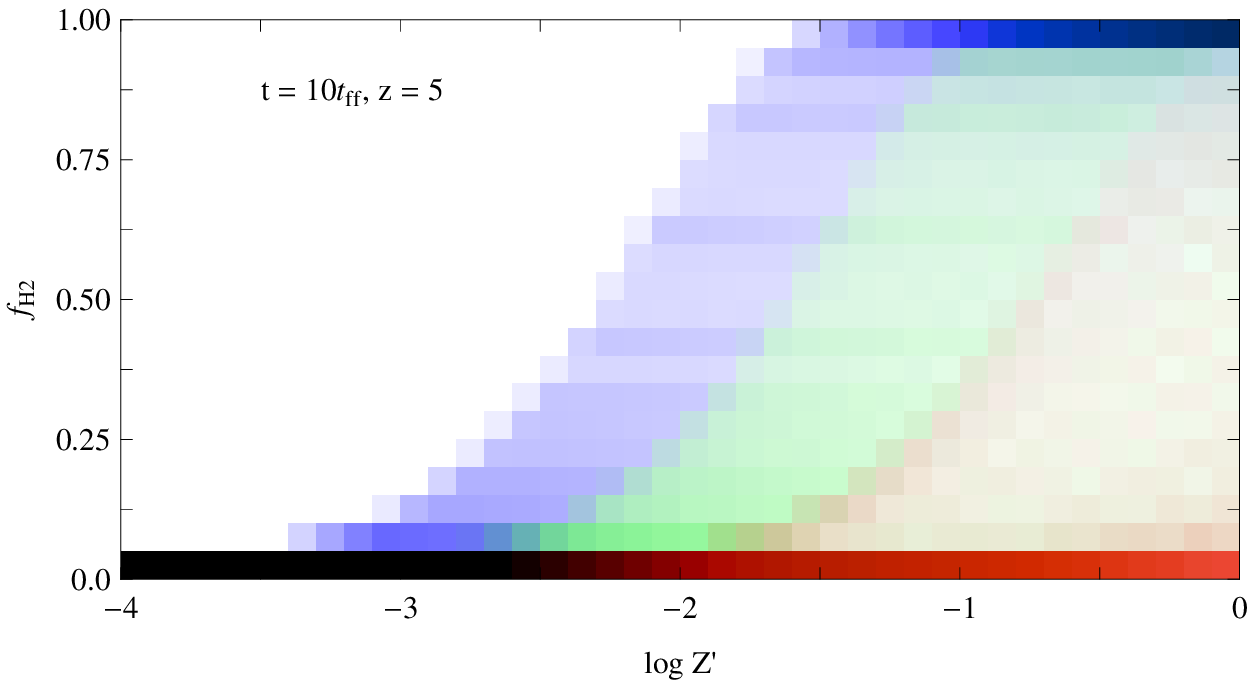}

\plotone{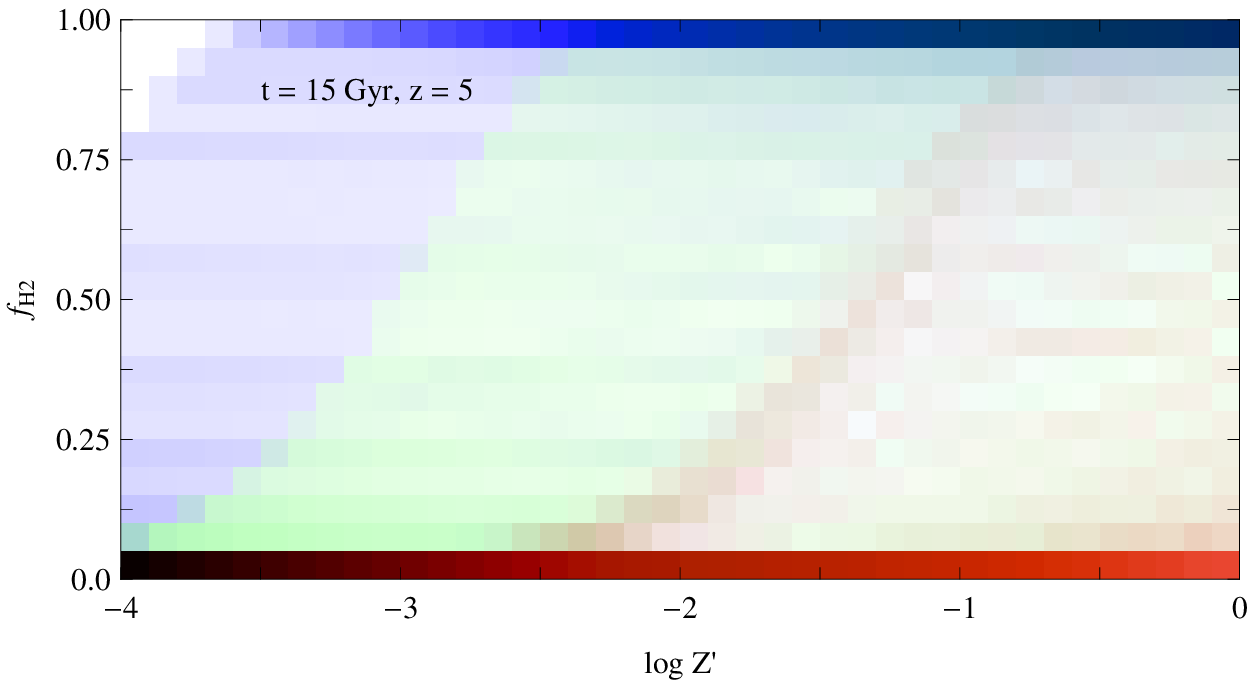}
\caption{
\label{fig:fH2ZgridCMBz5}
Same as Figure \ref{fig:fH2Zgrid}, but computed using a minimum gas temperature equal to the CMB temperature $T_{\rm CMB} = 2.73(1+z)$ K at $z=5$.
}
\end{figure}

\begin{figure}
\plotone{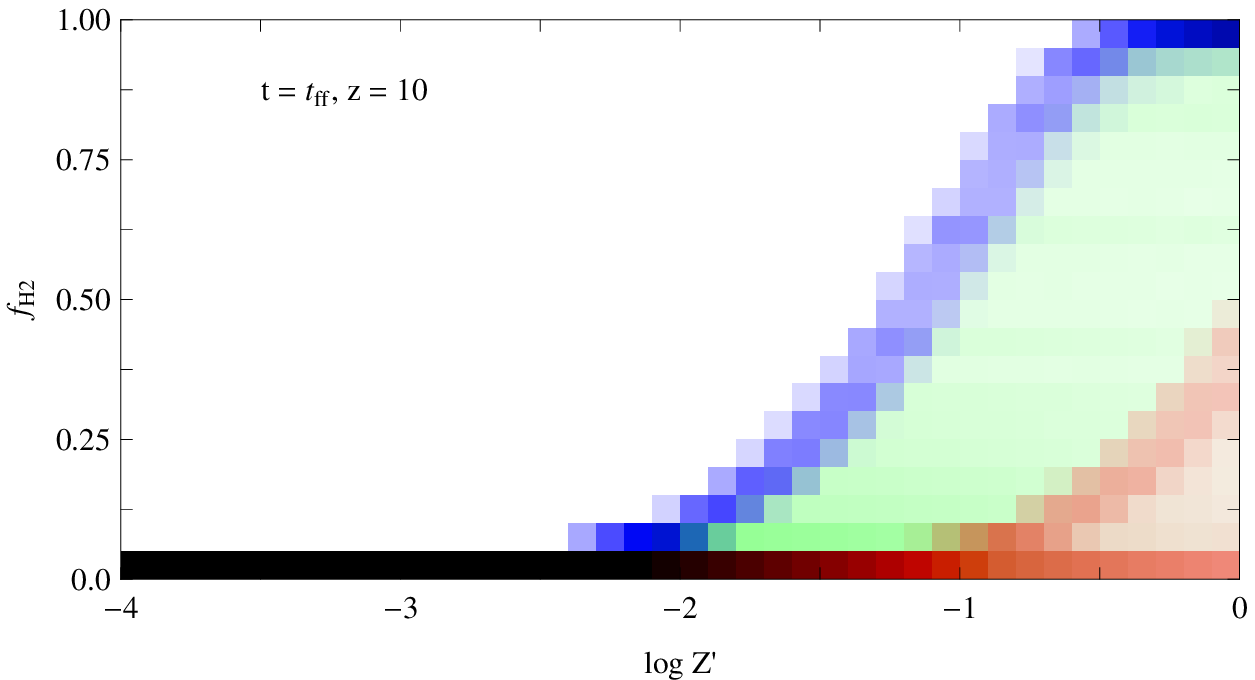}

\plotone{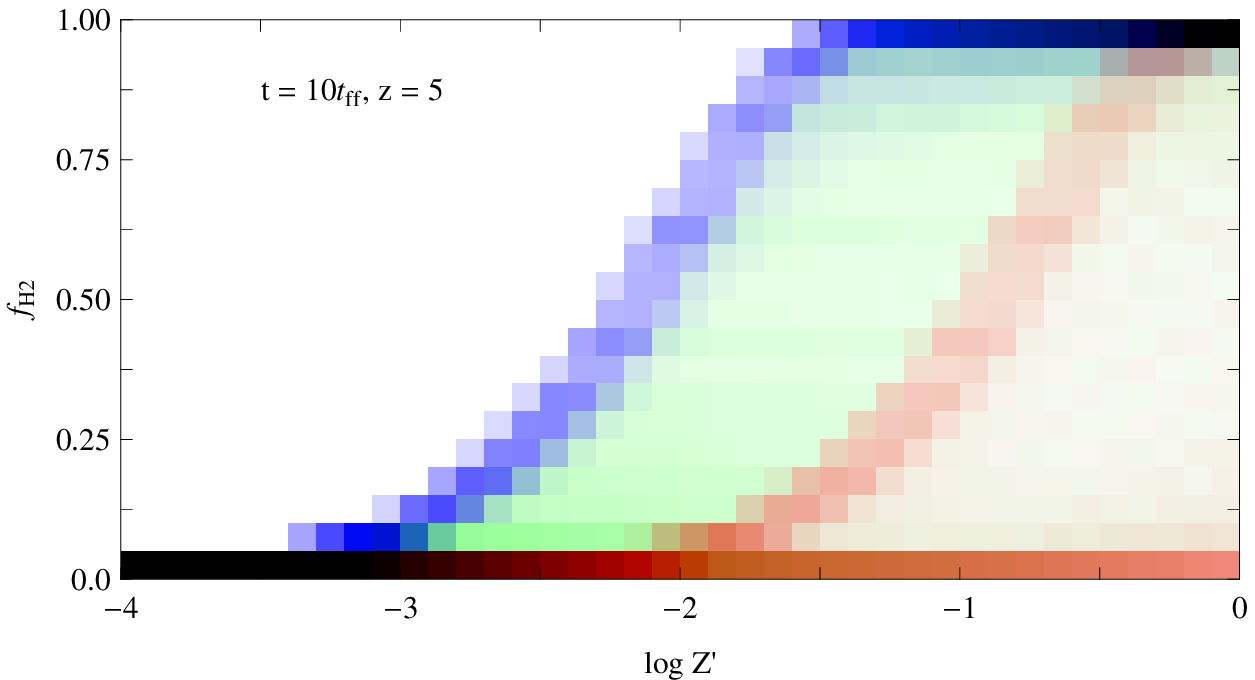}

\plotone{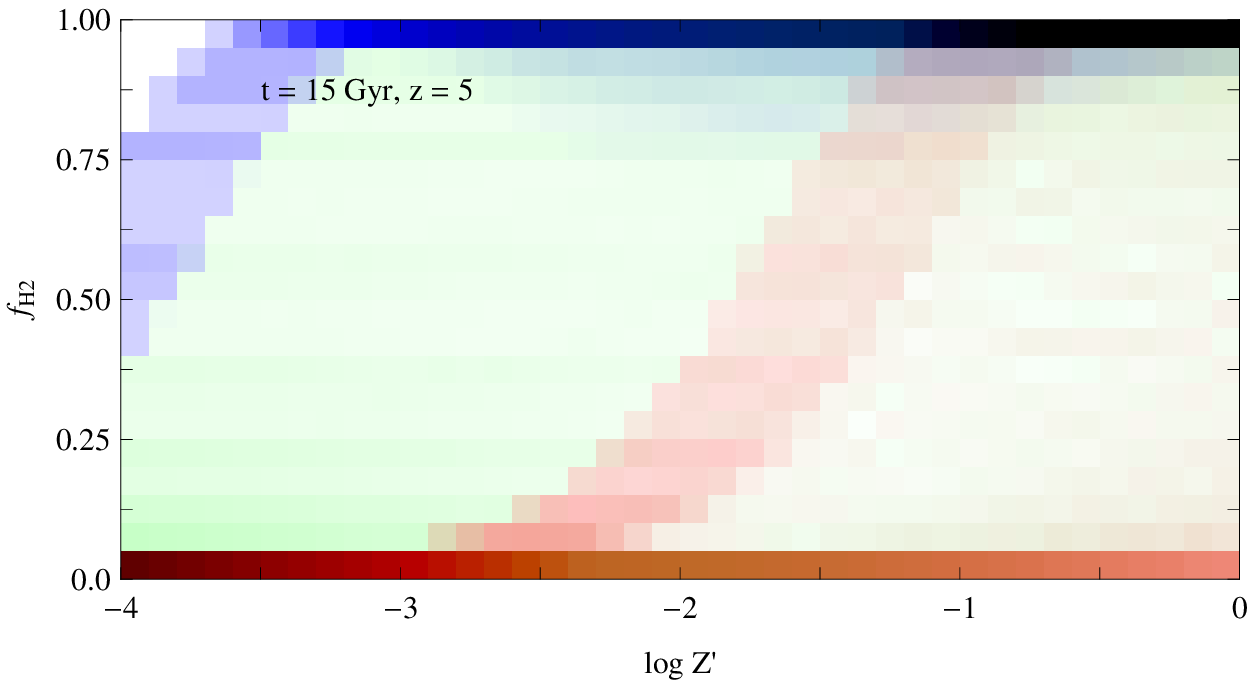}
\caption{
\label{fig:fH2ZgridCMBz10}
Same as Figure \ref{fig:fH2Zgrid}, but computed using a minimum gas temperature equal to the CMB temperature $T_{\rm CMB} = 2.73(1+z)$ K at $z=10$.
}
\end{figure}

To determine the sensitivity of these results to the choices of radiation field and cosmic ray intensity, I also compute the grid with a radiation field increased by a factor of 30 compared to the Milky Way value ($\Gamma_{\rm PE,0} = 1.2\times 10^{-25}$ erg s$^{-1}$ and $\zeta_{\rm diss,0} = 1.5\times 10^{-10}$ s$^{-1}$) and with a cosmic ray flux increased by a factor of 10 to match the observed diffuse cloud value ($\zeta_{\rm CR} = 2\times 10^{-17}Z'$ s$^{-1}$). The results are shown in Figures \ref{fig:fH2Zgridhi} and \ref{fig:fH2Zgridhicr}. Comparison with Figure \ref{fig:fH2Zgrid} clearly indicates that the qualitative results are not substantially altered.

Finally, note that in the thermal evolution calculation I have neglected heating due to cosmic microwave background (CMB) photons. These will impose a temperature floor $T=2.73 (1+z)$ K, where $z$ is the redshift. A priori one would not expect the CMB to become significant until very high redshifts. The temperature reached by C~\textsc{ii} cooling does not fall below $\sim 20$ K over most of the model grid, and the CMB temperature does not exceed this value until $z>6.3$. To confirm this intuition, in Figures \ref{fig:fH2ZgridCMBz5} and \ref{fig:fH2ZgridCMBz10} I show the results of imposing a minimum temperature $T=2.73 (1+z)$ K on the temperature used to evaluate $M_{\rm BE}$, for $z = 5$ and $z = 10$. The changes in the results from Figure \ref{fig:fH2Zgrid} are essentially invisible at $z=5$. At $z=10$, the higher CMB temperature raises the temperature in some models such that there are fewer models with small values of $M_{\rm BE}$, and these cluster at even higher molecular fractions. Qualitatively, however, the results are the same as at lower $z$.

\section{Discussion}

\subsection{Observational Implications and Tests}

\begin{figure}
\plotone{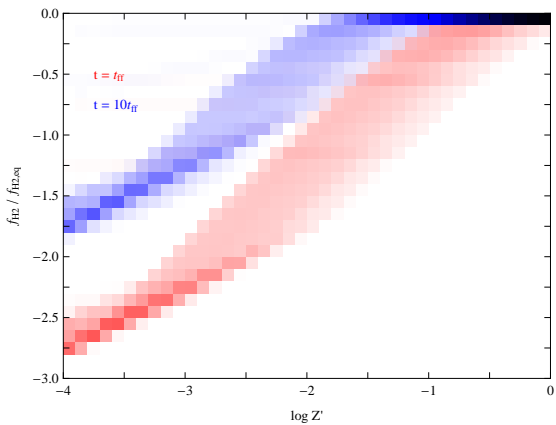}
\caption{
\label{fig:h2ratio}
Ratio of H$_2$ fraction at time $t=t_{\rm ff}$ (red) and $t=10t_{\rm ff}$ (blue) to the equilibrium value. For each model in a grid covering the parameter range $\log Z' = -4 - 0$, $\log n_0 = 0 - 3$, $\log A_V = -2 - 1$, I compute the Bonnor-Ebert mass $M_{\rm BE}$ H$_2$ fraction at times $t = t_{\rm ff}$ and $t=10 t_{\rm ff}$. For this plot, I retain only models with $M_{\rm BE}<100$ $\msun$ at $t=t_{\rm ff}$, indicating that these clouds are likely to form stars. As in Figure \ref{fig:fH2Zgrid}, pixel brightness indicates what fraction of the models at a given $Z'$ fall into that particular bin of $f_{\rm H_2} / f_{\rm H_{2,\rm eq}}$, with white indicating no models and solid red or blue indicating all the models.
}
\end{figure}

\begin{figure}
\plotone{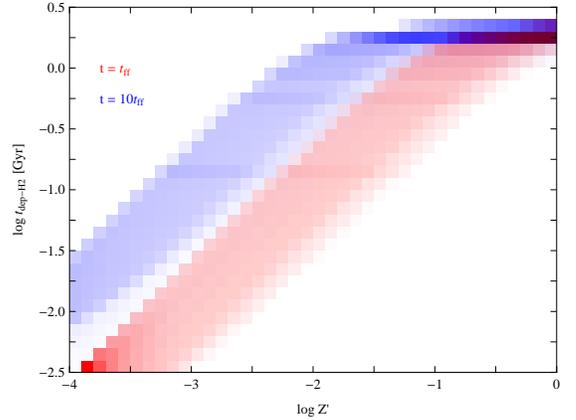}
\caption{
\label{fig:tdep}
Same as Figure \ref{fig:h2ratio}, except that the $y$ axis now represents the H$_2$ depletion time $t_{\rm dep-H_2} = f_{\rm H_2} (2\mbox{ Gyr})$.
}
\end{figure}

The disconnect in timescales between H$_2$ formation and cooling has two major observable consequences, which can be used as a test of the above calculations. The first of these is a drop in H$_2$ fractions below the levels predicted by equilibrium models in star-forming clouds. Figure \ref{fig:h2ratio} shows the ratio of the H$_2$ fraction at $t=t_{\rm ff}$ and $t=10t_{\rm ff}$ to the equilibrium H$_2$ fraction for star-forming clouds in the model grid. Clearly we expect equilibrium models to provide good predictions for galaxies down to metallicity $Z' \approx 0.1$ or even somewhat less. This is consistent with observations to date, which show that chemical equilibrium models provide excellent fits to observed H$_2$ to H~\textsc{i} ratios in the Milky Way \citep{krumholz09a, lee12a} and even the Small Magellanic Cloud (SMC; $Z'\approx 0.2$, \citealt{bolatto11a}). However, the Figure indicates that at metallicities of $\log Z' = -3$, the H$_2$ fraction in a given star-forming cloud will be at most $\sim 10\%$ of its equilibrium value, and could be less than 1\% of that value.

Second, the onset of star formation before the gas has time to fully transform to H$_2$ in low metallicity galaxies should manifest as a reduction in the H$_2$ depletion time $t_{\rm dep-H_2}$, defined as the ratio of the H$_2$ mass to the star formation rate. In Solar metallicity, non-starbursting local galaxies, $t_{\rm dep-H_2} \approx 2$ Gyr \citep{bigiel08a}, although lower values are possible in starbursts. This value should be lower in low metallicity galaxies by a factor of the mean H$_2$ fraction in cold, star-forming clouds, since these clouds will only partially convert to H$_2$ before forming stars and being destroyed by feedback. Figure \ref{fig:tdep} illustrates this effect for clouds that live 1 and 10 free-fall times. Note that \citet{glover12b} qualitatively suggested the existence of this effect, and Figure \ref{fig:tdep} represents a quantitative extension of this prediction.

Observational tests of these predictions are complicated by the fact that H$_2$ is extremely difficult to observe at low metallicities, because CO, the traditional H$_2$ proxy, ceases to track H$_2$ at metallicities below a few tenth of Solar \citep{krumholz11b, bolatto11a, leroy11a, shetty11b, narayanan12a, feldmann12a}. Thus direct observational tests will require the detection of H$_2$ by other means, such as dust or C~\textsc{ii} emission that is not associated with observed H~\textsc{i}. While observationally challenging, surveys of this sort have already been completed in the closest galaxies like the SMC \citep{bolatto11a}, and with the observational power provided by the Atacama Large Millimeter Telescope (ALMA) should begin to be possible in even lower metallicity nearby galaxies. In particular, 850 $\mu$m observations are an excellent probe of dust and thus all gas including H$_2$, because at 850 $\mu$m dust is generally optically thin, the emission is not very sensitive to dust temperature, and ALMA can achieve both high spatial resolution and excellent sensitivity. Prime targets for such a campaign include IZw18, SBS 0335-052 (both $Z' \approx 0.02$, and probably even lower dust metallicities, \citealt{izotov99a, herrera-camus12a}), and Leo T ($Z'\ltsim 0.01$, \citealt{simon07a}). The ALMA observations will have to be coupled with high resolution, high sensitivity H~\textsc{i} maps to measure the atomic content. Fortunately, sub-kpc resolution H~\textsc{i} maps of IZw18 \citep{van-zee98a} and SBS0335-052 \citep{ekta09a} are already available in the literature.

\subsection{Implications for Simulations and Semi-Analytic Models}

These results have important theoretical implications as well. Many galaxy simulation models allow star formation only in regions where the gas has converted to H$_2$; some of these models include non-equilibrium chemistry for H$_2$ formation and destruction \citep{pelupessy09a, gnedin09a, gnedin10a, christensen12a}, while others assume equilibrium \citep{fu10a, lagos11a, kuhlen12a, krumholz12d}. The non-equilibrium models on average yield less H$_2$ and thus less star formation at low metallicity, because often gas clouds are not able to build up significant H$_2$ fractions before being destroyed by galactic shear or similar kinematic processes \citep{krumholz11a}. However, if the relevant timescale is the cooling time and not the H$_2$ formation time, and this effect should be far less significant. As a result, star formation should in fact occur even in gas with low H$_2$ fractions, provided that the {\it equilibrium} H$_2$ fraction is high -- it is the equilibrium H$_2$ fraction and not the instantaneous one that correlates with gas temperature and thus is a good predictor of where star formation will occur. This suggests that, ironically, models in which the H$_2$ is assumed to be in equilibrium, while they are less accurate in predicting the actual H$_2$ fraction, may in fact be more accurate that the non-equilibrium models in predicting where star formation should occur. More generally, the calculations presented here suggest that star formation thresholds in simulations should be based on the instantaneous density and extinction, which determine the temperature, and not on non-equilibrium chemical abundances.

\section{Summary}

I explore under what conditions and for what physical reasons the observed correlation between star formation and molecular gas in the ISM is likely to break down. I show that the breakdown occurs at metallicities below a few percent of Solar, and that the physical mechanism for this breakdown is a disconnect between the thermal and chemical equilibration timescales. Carbon in the ISM is able to cool gas on a timescale shorter by a factor of several thousand than that required for dust grains to convert the H~\textsc{i} to H$_2$. As long as both the thermal and chemical equilibration timescales are short compared to cloud free-fall times, which is the case at Solar metallicity, this does not have any practical effect and non-equilibium chemistry is unimportant. However, both the thermal and chemical timescales scale linearly with the metallicity, while the free-fall time time does not. At metallicities below a few percent of Solar, the free-fall time becomes intermediate between the thermal and chemical timescales, and clouds cool and proceed to star formation before molecules form, breaking the H$_2$-star formation connection.

This result has three major implications, two observational and one theoretical. The observational implications are that the equilibrium chemistry models that perform extremely well in the Milky Way and the SMC should begin to overpredict H$_2$ abundances in very low metallicity galaxies, and that star formation should occur in atomic-dominated regions of such galaxies as well, leading to a lower H$_2$ depletion time. These predictions are not trivial to check, given the difficulty of measuring H$_2$ in low metallicity environments, but combining high resolution dust and H~\textsc{i} maps to infer the presence of H$_2$ constitutes a viable strategy. The theoretical implication is that galaxy evolution simulations and semi-analytic models that link star formation to the chemical state of the gas, and that treat that chemistry using non-equilibrium models, are likely to underpredict star formation rates in circumstances where the gas should reach thermal but not chemical equilibrium. It is the former that matters for star formation, not the latter.

\acknowledgements I thank A.~Bolatto, L.~Hunt, and A.~Leroy for helpful conversations, and C.~McKee for helpful comments on the manuscript. Funding for this work as provided by the National Science Foundation through grant CAREER-0955300, by NASA through Astrophysics Theory and Fundamental Physics Grant NNX09AK31G and through a Chandra Space Telescope Grant, and by the Alfred P.~Sloan Foundation.


\end{document}